\RequirePackage{ifpdf}
\ifpdf 
\documentclass[pdftex]{sigma}
\else
\documentclass{sigma}
\fi

\begin{document}

\allowdisplaybreaks

\renewcommand{\PaperNumber}{001}

\FirstPageHeading

\renewcommand{\thefootnote}{$\star$}

\ShortArticleName{Normalization Factors and Crypto-Hermitian Quantum
Models}

\ArticleName{On the Role of the  Normalization Factors $\boldsymbol{\kappa_n}$\\
and of the Pseudo-Metric $\boldsymbol{{\cal P} \neq {\cal P}^\dagger}$\\ in Crypto-Hermitian Quantum
Models\footnote{This paper is a
contribution to the Proceedings of the 3-rd Microconference
``Analytic and Algebraic Me\-thods~III''. The full collection is
available at
\href{http://www.emis.de/journals/SIGMA/Prague2007.html}{http://www.emis.de/journals/SIGMA/Prague2007.html}}}

\Author{Miloslav ZNOJIL}

\AuthorNameForHeading{M. Znojil}

\Address{\'{U}stav jadern\'e fyziky AV \v{C}R, 250 68
\v{R}e\v{z}, Czech Republic}
\Email{\href{mailto:znojil@ujf.cas.cz}{znojil@ujf.cas.cz}}
\URLaddress{\url{http://gemma.ujf.cas.cz/~znojil/}}

\ArticleDates{Received November 26, 2007; Published online January 02, 2008}

\Abstract{Among ${\cal P}$-pseudo-Hermitian Hamiltonians $H ={\cal
P}^{-1}\,H^\dagger\,{\cal P}$ with real spectra, the ``weakly
pseudo-Hermitian" ones (i.e., those employing non-self-adjoint
${\cal P} \neq {\cal P}^\dagger$)  form a remarkable subfamily. We
list some reasons why it deserves a special attention. In
particular we show that whenever ${\cal P} \neq {\cal P}^\dagger$,
the current involutive operator of charge ${\cal C}$ gets
complemented by a nonequivalent alternative involutive quasiparity
operator ${\cal Q}$. We show how, in this language, the standard
quantum mechanics is restored via the two alternative inner
products  in the physical Hilbert space of states, with $\langle
\psi_1\,|\,{\cal PQ}\,|\,\psi_2\rangle=\langle \psi_1\, | \, {\cal
CP}\,|\,\psi_2\rangle $.}


\Classification{81Q10; 47B50}

\renewcommand{\thefootnote}{\arabic{footnote}}
\setcounter{footnote}{0}

\section{Introduction}

One of the most important keys to the solvability of
Schr\"{o}dinger equations
 \[
 H\,|\,\psi\rangle = E\,|\,\psi\rangle
 \]
is often found in the existence of a symmetry $S$ of the
Hamiltonian $H$, i.e., in the commutativity
 \begin{gather}
 HS-SH=0.
 \label{comm}
 \end{gather}
During the development of Quantum Mechanics, the concept of
symmetry found various genera\-lizations. For illustration, one
could recollect the multiple applications of Lie algebras (where~$H$ appears as just one of their generators) or supersymmetries
(where one employs both the commutators and anticommutators).

Recently \cite{BB}, the family of the productive symmetry-related
mathematical tools has been enriched by the so called ${\cal
PT}$-symmetry where the vanishing commutator (\ref{comm}) contains
an {\em  antilinear} operator $S={\cal PT}$
\cite{ali,Solombrino,Solombrinob}. In the context of f\/ield theory,
typically, ${\cal P}$ is chosen as parity while the antilinear
Hermitian-conjugation factor ${\cal T}$ mimics time reversal
\cite{BB,BBb,Carl}. More rigorously~\cite{ali}, one replaces
equation~(\ref{comm}) by the requirement
 \begin{gather}
 H^\dagger = {\cal P}\,H\,{\cal P}^{-1} .
 \label{ptsymm}
 \end{gather}
In an illustrative two-by-two matrix model with
 \begin{gather}
 H^{(2)}=
 \left (
 \begin{array}{cc}
 a&b\\ c&d
 \end{array}
 \right ),\quad a,b,c,d \in \mathbb{C},\qquad
 {\cal P} =
 \left (
 \begin{array}{cc}
 1&0\\ 0&-1
 \end{array}
 \right )
 \label{tri}
\end{gather}
we get the three constraints
 $ a=a^*$, $d=d^*$, $b=-c^* $.
Our ${\cal PT}$-symmetric toy Hamiltonian~$H^{(2)}$ has four free
real parameters (same number as if it were Hermitian) and its four
energies~$E$ remain real (i.e., in principle, observable) in a
specif\/ic ``physical'' subdomain ${\cal D}$ of its matrix elements
where $(a-d)^2\geq 4bb^*$. As long as there would be no such a
constraint in Hermitian case, new interesting physical as well as
mathematical phenomena can be expected to occur along the
``exceptional-point'' \cite{Kato} boundary $\partial {\cal D}$
where $2|b|=|a-d|$.

Inside ${\cal D}$, in the light of the review paper \cite{Geyer},
the model $H^{(2)}$ should be called ``quasi-Hermitian" since, by
construction,  all its spectrum is real. This means that our
matrix $H^{(2)}$ becomes Hermitian in the (two-dimensional) vector
space where the scalar product between ele\-ments $|\,a\rangle$ and
$|\,b\rangle$ is def\/ined by the overlap $\langle
a\,|\,\Theta\,|\,b\rangle$ where $\Theta=\Theta^\dagger>0$ is a
suitable matrix solution of the quasi-Hermiticity condition of~\cite{Geyer},
 \begin{gather}
 H^\dagger = \Theta\,H\,\Theta^{-1}.
 \label{cptsymm}
\end{gather}
{\it Mutatis mutandis}, all these considerations can be easily
transferred to an arbitrary inf\/inite-di\-men\-sional Hilbert space
${\cal H}$ where the Hamiltonians $H$ have to be assigned a {\em
positive-definite\/} ope\-rator $\Theta=\Theta^\dagger$ exhibiting all
the necessary mathematical properties of the metric in ${\cal
H}={\cal H}^{(\Theta)}$~\cite{Geyer}. Thus, the usual {\em single}
standard Hermiticity condition $H=H^\dagger$ is replaced  by the
{\em pair} of the gene\-ralized symmetry rules (\ref{ptsymm}) and
(\ref{cptsymm}). Also the concept and construction of observables
becomes perceivably modif\/ied. This def\/initely opens new horizons
in quantum phenomenology~\cite{Carl}.

In the related literature (we recommend its long list collected in
\cite{Carl}), it is not always suf\/f\/iciently emphasized that the
proper physical meaning of equations~(\ref{ptsymm}) and (\ref{cptsymm})
is in fact perceivably dif\/ferent. Indeed, the latter,
quasi-Hermiticity condition (\ref{cptsymm}) is ``strong'' (it
guarantees that $E$s are real) and
``dif\/f\/icult''\footnote{Technically, the construction of $\Theta$ is
almost never easy; in the case of our present two-by-two example,
\cite{Hendrik}~could be consulted for an explicit
illustration of the form of $\Theta$ etc.}. In contrast, the former
condition (\ref{ptsymm}) (called, usually, ${\cal P}$-pseudo-Hermiticity \cite{ali}) is just auxiliary (in fact, we
need it just for certain technical purposes~-- see below) and
``almost redundant''\footnote{It is, in fact, neither necessary nor
suf\/f\/icient for the reality of the energies; sometimes, the
concrete choice of~${\cal P}$ is even left unspecif\/ied
\cite{ali}.}.

From a historical point of view it is a paradox that in spite of
the knowledge of the aspects and merits of equation~(\ref{cptsymm})
(cf.~\cite{Geyer} with examples from nuclear physics), it
was just the ``naive'' parity-pseudo-Hermiticity property
(\ref{ptsymm}) of certain models which proved much more inspiring.
Anyhow, several aspects of its formal appeal (thoroughly listed in~
\cite{Carl}) attracted attention to the whole new class of the
models which were often neglected in the past because they
happened to be non-Hermitian with respect to the ``Dirac's'' very
special metric $\Theta^{\rm (Dirac)}=I$.

\section{Pseudo-Hermitian Hamiltonians }

A broad menu of the new, quasi-Hermitian\footnote{Or, in a
terminology coined recently by Andrei Smilga \cite{Smilga},
``crypto-Hermitian''.} models has been studied in the literature
after their pioneering sample has been of\/fered by Bender and
Boettcher in 1998 \cite{BB,BBb}. Among them, paradoxically, some
of the most important ones were {\em not} using the parity
operator (\ref{tri}) but rather its two-by-two basis-permutation
alternative
 \begin{gather}
 {\cal P} =
 \left (
 \begin{array}{cc}
 0&1\\1&0
 \end{array}
 \right ).
 \label{psptsymm}
 \end{gather}
The related modif\/ication of equation~(\ref{ptsymm}) is encountered not
only in the well known Feshbach's and Villar's version of the
Klein--Gordon equation describing relativistic spinless bosons
\cite{FV,jaKG} but also in certain equations employed in quantum
cosmology~\cite{aliKG}, in non-Hermitian but ${\cal PT}$-symmetric
coupled-channel problems~\cite{cc} and, unexpectedly, even in
classical magnetohydrodynamics~\cite{Uwe} and electrodynamics~\cite{elek}. For us, the unexpected and surprisingly widespread
applicability of models based on the basis-permutation matrix
structure (\ref{psptsymm}) of ${\cal P}$ in (\ref{ptsymm})
provided a strong support of our continuing interest in the more
complicated non-parity generalizations of the Hermitian
pseudometrics\footnote{You could also call it parity, in broader
sense.} ${\cal P}={\cal P}^\dagger$ \cite{regukl}.

In the next step of our study of the models with
 $H \neq H^\dagger$
we were led to an active interest in the weakly
pseudo-Hermiticitian cases (introduced by Solombrino
\cite{Solombrino}) where {\em non-Hermitian} pseudoparities
 ${\cal P} \neq {\cal P}^\dagger$
are admitted. In particular, we contemplated the ``f\/irst
nontrivial'' three-dimensional basis-permutations ${\cal P}$ of the
non-Hermitian form in~\cite{str}. To our greatest surprise we
revealed that the family of the expectedly more f\/lexible
three-dimensional descendants of the above two-dimensional model
(\ref{tri}), viz.,
 \begin{gather}
 {\cal P} =\left (
 \begin{array}{ccc}
 0&0&1\\
 1&0&0\\
 0&1&0
 \end{array}
 \right )=\left ({\cal P}^\dagger \right )^{-1}=
 {\cal P}^{-2},\qquad
 H^{(3)}= \left (
 \begin{array}{ccc}
 a&b&b^*\\
 b^*&a&b\\
 b&b^*&a
 \end{array}
 \right )
 \label{provazanekanaly3}
 \end{gather}
is in fact much more constrained as it contains {just three} free
real parameters again, with $a,b \in \mathbb{C}$ but $a=a^*$. For
this reason we proposed to rename its ``weak'' pseudo-Hermiticity
feature into a ``strengthened ${\cal PT}$-symmetry''.

Recently, Ali Mostafazadeh re-focused attention on our
model~(\ref{provazanekanaly3}) in \cite{aliweak}, emphasizing that
the weak pseudo-Hermiticity and pseudo-Hermiticity specify the
same class of operators in f\/inite dimensions (cf.\ also
\cite{well}). As an illustrative example he recalls our
equation~(\ref{provazanekanaly3}) and argues that our model $H^{(3)}$
proves {\em also} pseudo-Hermitian with respect to the {\it
Hermitian}
 \begin{gather}
 {\cal P}^{(+)}={\cal P}+{\cal P}^\dagger.
 \label{BQP}
 \end{gather}
A similar remark could have been also deduced from the older
comment \cite{BQ} by Bagchi and Quesne who, apparently, did not
notice that the trick might lead to a {\em singular} and, hence,
unacceptable~${\cal P}^{(+)}$ in general. In this sense one should
appreciate that A.~Mostafazadeh~\cite{aliweak} found an elegant
way out of the trap by the mapping of a given
 ${\cal P} \neq {\cal P}^\dagger$
on the whole one-parametric set of its eligible Hermitian partners
 \begin{gather}
  {\cal P}^{(AM)}(\theta)=
 {\rm i}\big [{\cal P} \exp ({\rm i}\theta)
 -{\cal P}^\dagger \exp(- {\rm i}\theta)\big ]
 \label{subsum}
 \end{gather}
(given by equation~(19) of~\cite{aliweak}). In what follows we
intend to add several further remarks on the specif\/ic character
and merits of all the models $H$ which are characterized by such
an unexpectedly  large freedom in the choice between the
alternative ``pseudo-metrics'' given by equation~(\ref{subsum}).

First of all, we would like to point out that from the purely
pragmatic point of view there is an obvious dif\/ference between the
strongly constrained {\em three-parametric} $H^{(3)}$-toy-model
subfamily~(\ref{provazanekanaly3}) and the {\em much
broader}\footnote{In fact, seven-parametric.} class of the generic
${\cal P}^{(+)}$-pseudo-Hermitian three-dimensional Hamiltonians.
In this sense, equation (\ref{subsum}) only enters the scene as a
natural complement and extension of equation~(\ref{BQP}) and as a very
useful tool of a subsumption of some families of the Hamiltonians
(in this sense one only has to get accustomed to the fact of life
that inside the family of the $N$-dimensional pseudo-Hermitian
$H$s there exists just a very small ``weakly''-pseudo-Hermitian
subfamily).

Secondly, let us add that the f\/lexible recipe (\ref{subsum}) would
f\/ind its applicability in the general $N$-dimensional context of
our systematic coupled-channel study \cite{asym} where the
Hermitian partner of the $N=4$ pseudo-metric
 \[
 {\cal {P}} =\left (
 \begin{array}{cccc}
 0&0&0&1\\
 1&0&0&0\\
 0&1&0&0\\
 0&0&1&0
 \end{array}
 \right )\neq {\cal {P}}^\dagger=
 {\cal {P}}^{-1}
 \]
would remain non-invertible whenever represented by the older
formula (\ref{BQP}), i.e., whenever $\theta$ in~(\ref{subsum})
were chosen as an integer multiple of $\pi/2$.

Thirdly, let us emphasize that in physics, the {\em only
essential} feature of the Hamiltonians $H\neq H^\dagger$ is in
fact represented by their quasi-Hermiticity property~(\ref{cptsymm}). It is clear that the pseudo-Hermiticity itself is
much less relevant because once we get through the {\em difficult}
proof of the necessary reality of the spectrum \cite{BBb}, the
pseudo-Hermiticity of a given $H$ becomes in fact {\em fully
equivalent} to its quasi-Hermiticity \cite{ali,Solombrinob}. In
this context we are sure that a more explicit evaluation of some
additional practical dif\/ferences between the more or less {purely
technical} assumptions ${\cal P}= {\cal P}^\dagger$ and ${\cal
P}\neq {\cal P}^\dagger$ could enhance our understanding of the
specif\/ic merits of certain specif\/ic choices of the non-Hermitian
models $H$ with real spectra.

\section[Metrics $\Theta$]{Metrics $\boldsymbol{\Theta}$}

We are now going to propose a possible comparison between the
pseudo-Hermiticity (sampled by equation~(\ref{ptsymm}) where
 ${\cal P}= {\cal P}^\dagger$)
and the weak  pseudo-Hermiticity (sampled by equation~(\ref{ptsymm})
where ${\cal P}\neq {\cal P}^\dagger$). Our main idea is twofold.
Firstly, we recollect that the simpler the ${\cal P}$, the simpler
are the explicit formulae for the basis (cf.\ Subsection~\ref{3a}
below). Secondly, in Subsections~\ref{3b} and~\ref{3c} we shall
draw some consequences from the fact that in the majority of
applications of non-Hermitian Hamiltonians, the most important
role played by~${\cal P}$ is its occurrence  in the factorized
metric~$\Theta$~\cite{Carl,pseudo}.

For {\em any} given observable ${\cal O}$, the knowledge of the
metric is essential for the practical evalua\-tion of its (real)
expectation values
 \[
 \langle\,\psi\,|\Theta\,{\cal O}\,|\,\psi\rangle.
 \]
The quantum system can be prepared in a complicated state
$|\psi\rangle \in {\cal H}^{(\Theta)}$ so that the factorization
 $\Theta={\cal CP}$
can be of a key technical signif\/icance. It is equally important
that this factorization enables us to formulate an important
additional postulate ${\cal C}^2=I$ which is often deeply rooted
in certain hypothetical physics considerations~\cite{Carl}. Even
on a purely formal level, the latter postulate represents one of
the most widely accepted ways of getting rid of the well known and
highly unpleasant  ambiguity \cite{Geyer,Hendrik,pseudo} of the
general solutions $\Theta$ of the quasi-Hermiticity constraint~(\ref{cptsymm}).

Once we turn our attention to the models where
 ${\cal P}\neq {\cal P}^\dagger$,
their {\em different} nature becomes obvious once we interpret
them as resulting from an application of a symmetry of the generic
form (\ref{comm}). We arrive at the f\/irst specif\/ic feature of the
weak pseudo-Hermiticity which, strictly speaking, replaces
equation~(\ref{comm}) (containing a {\em single} antilinear operator
$S={\cal PT}$) by the {\em triplet of parallel requirements}
 \begin{gather*}
 H^\dagger = {\cal P} H {\cal P}^{-1},
 \qquad
 H^\dagger = {\cal P}^\dagger H
 \left [{\cal P}^{-1}
 \right ]^\dagger,
 \qquad
 [H ,S]=0,\qquad
 S={\cal P}^{-1}{\cal P}^\dagger.
 \end{gather*}
Although just two of them are independent of course, we already
illustrated how they impose {\em much more stringent} constraints
upon $H$.

\subsection{The family of biorthogonal bases}\label{3a}

In order to proceed to the technical core of our present message
let us f\/irst stay in the ``usual'', auxiliary and non-physical
Hilbert space ${\cal H}^{(I)}$ and treat a given $H\neq H^\dagger$
with real spectrum~$\{E_n\}$ as ``non-Hermitian''. Using a slightly
modif\/ied Dirac's notation we may f\/ind the respective left and
right eigenvectors $|\,E_n\rangle^{(1)}$ and
$^{(1)}\langle\!\langle E\,|$ of our $H$ from the corresponding
doublet of Schr\"{o}dinger equations,
 \begin{gather}
 H\,|\,E_n\rangle^{(1)}=E_n\,|\,E_n\rangle^{(1)},
 \label{cerka}
 \\
 ^{(1)}\langle\!\langle E_m\,|\,H= E_m\,^{(1)}\langle\!\langle E_m\,| .
 \label{derka}
 \end{gather}
The reason for our introduction of a superscript $^{(1)}$ lies in
the fact that even if we impose the standard biorthonormality
conditions
 \[
 ^{(1)}\langle\!\langle E_m\,|\,E_n\rangle^{(1)} =\delta_{mn}
 \]
accompanied by the standard completeness formula in ${\cal
H}^{(I)}$,
 \[
 \sum_{n=0}^\infty\,|\,E_n\rangle^{(1)}\,^{(1)}\langle\!\langle E_n\,|
  =I
 \]
we can still redef\/ine our eigenvectors by the formula
 \begin{gather}
 ^{(\vec{\kappa})}\!\langle\!\langle E_m\,|
 =
 ^{(1)}\!\!\langle\!\langle E_m\,|\cdot \frac{1}{\kappa_n}\,,
\nonumber \\
 |\,E_n\rangle^{(\vec{\kappa})} =
 |\,E_n\rangle^{(1)}\cdot \kappa_n
 \label{renorm}
 \end{gather}
with arbitrary complex $\kappa_0 ,  \kappa_1 ,  \kappa_3 ,
 \ldots$ forming an inf\/inite-dimensional vector $\vec{\kappa}$.

 \section{The role of the set of the normalization factors}

We saw that once we change any the normalization constant
$\kappa_n$ we arrive at another, ``renormalized" biorthonormal set
exhibiting the {\em same} eigenenergy, orthonormality and
completeness properties. Obviously, the  freedom of this type
would vanish completely whenever one returns to the Hermitian
Hamiltonian operators $H$. In an opposite direction, the specif\/ic
relevance of the variability of the normalization factors
$\kappa_n$ becomes more important in the scenarios where ${\cal P}
\neq {\cal P}^\dagger$.

In the characteristic latter case one assumes that ${\cal P}$
remains extremely elementary. For this reason, even the transition
to the Hermitian pseudometric (\ref{subsum}) could make some
formulae much less transparent. In what follows, we intend to
describe a particularly interesting application of such an idea to
the specif\/ic, very popular models where one constructs the metric
$\Theta$ in a~factorized form.

\subsection[The operators ${\cal Q}$ of quasiparity]{The operators $\boldsymbol{\cal Q}$ of quasiparity}\label{3b}

In the generic non-degenerate case with
 $H^\dagger = {\cal P} H {\cal P}^{-1}$
and with the non-Hermitian ${\cal P} \neq {\cal P}^\dagger$, the
$^{(\vec{\kappa})}$-superscripted versions of equations~(\ref{cerka})
and (Hermitian conjugate) (\ref{derka}),
 \begin{gather*}
 H\,|\,E_n\rangle^{(\vec{\kappa})}=E_n\,|\,E_n\rangle^{(\vec{\kappa})},
 \\
 H^\dagger\,|\,E_n\rangle\!\rangle^{(\vec{\kappa})}=
 E_n\,|\,E_n\rangle\!\rangle^{(\vec{\kappa})}
 \end{gather*}
imply the proportionality of alternative solutions at the same
energy, say,
 \begin{gather}
 |\,E_n\rangle\!\rangle^{(\vec{\kappa})}=
 {\cal P}\,|\,E_n\rangle^{(\vec{\kappa})}\cdot q_n^{(\vec{\kappa})}  .
 \label{psiv}
 \end{gather}
As long as we normalized our basis at all $\vec{\kappa}$, we have
 \[
 1= ^{(\vec{\kappa})}\!\!\langle E_n\,|\,
 {\cal P}\,|\,E_n\rangle^{(\vec{\kappa})}\cdot q_n^{(\vec{\kappa})}
 \]
so that, in the light of equation~(\ref{renorm}), we have
$1=^{(1)}\!\!\langle E_n\,|\,
 {\cal P}\,|\,E_n\rangle^{(1)}\cdot
  q_n^{(\vec{\kappa})}\cdot
 \kappa_n^*\kappa_n$.
This leads to the renormalization-dependence formula
 \[
  q_n^{(\vec{\kappa})}= \frac{1}{\kappa_n^*\kappa_n}
  \,q_n^{(1)} , \qquad
  q_n^{(1)}=\frac{1}{^{(1)}\!\langle
E_n\,|\,
 {\cal P}\,|\,E_n\rangle^{(1)}}
 .
 \]
Now we may follow our old preprint~\cite{pseudo} and def\/ine the
{\em family} of the operators of quasiparity ${\cal Q}={\cal
Q}^{(\vec{\kappa})}$ by the relation
 \begin{gather}
 |\,E_n\rangle^{(\vec{\kappa})}
  \cdot q_n^{(\vec{\kappa})}=
  {\cal Q}^{(\vec{\kappa})}\,|\,E_n\rangle^{(\vec{\kappa})}
  \label{lviv}
 \end{gather}
inspired by equation~(\ref{psiv}) and leading to the spectral formula
with a simple manifest dependence on normalization,
 \[
 {\cal Q}^{(\vec{\kappa})}
 =\sum_{n=0}^\infty\,|\,E_n\rangle^{(\vec{\kappa})}
 \,{q_n^{(\vec{\kappa})}}\,
 ^{(\vec{\kappa})}\!\langle\!\langle E_n\,|=
 \sum_{n=0}^\infty\,|\,E_n\rangle^{(1)}
 \,\frac{q_n^{(1)}}{\kappa_n^*\kappa_n}\,
 ^{(1)}\!\langle\!\langle E_n\,| .
\]
We may conclude that equations~(\ref{psiv}) and (\ref{lviv}) lead to
the correct normalization recipe in the form
 \[
 ^{(\vec{\kappa})}\!\langle\,E_n\,|\,E_n\rangle\!\rangle^{(\vec{\kappa})}=
 ^{(\vec{\kappa})}\!\langle\,E_n\,|\,\Theta
 \,|\,E_n\rangle^{(\vec{\kappa})} =
 ^{(\vec{\kappa})}\!\langle\!\langle\,E_n\,|\,E_n\rangle^{(\vec{\kappa})}
 = 1
 \]
and, {\it ipso facto}, to the whole family
 \begin{gather}
 \Theta^{(\vec{\kappa})}={\cal P Q}^{(\vec{\kappa})}
 = \sum_{n=0}^\infty\,{\cal P}\,|\,E_n\rangle^{(1)}
 \,\frac{q_n^{(1)}}{\kappa_n^*\kappa_n}\,
 ^{(1)}\!\langle\!\langle E_n\,| \nonumber \\
\phantom{\Theta^{(\vec{\kappa})}}{} =\sum_{n=0}^\infty\,|\,E_n
 \rangle\!\rangle^{(1)}
 \,\frac{1}{\kappa_n^*\kappa_n}\,
 ^{(1)}\!\langle\!\langle E_n\,|=\sum_{n=0}^\infty\,|\,E_n
 \rangle\!\rangle^{((\vec{\kappa}))}
 \,\cdot\,
 ^{((\vec{\kappa}))}\!\langle\!\langle E_n\,|
 \label{sosonek}
  \end{gather}
of the manifestly renormalization-dependent and factorized,
self-adjoint, invertible and  positive def\/inite metric operators $
\Theta$.

\subsection[The operators ${\cal C}$ of charge]{The operators $\boldsymbol{\cal C}$ of charge} \label{3c}

In a close parallel to our preceding considerations we could have
also started from the Hermitian conjugate form of
equation~(\ref{cerka}) accompanied by the original equation~(\ref{derka}),
 \begin{gather*}
 ^{(\vec{\kappa})}\!\langle E_n\,|\,{\cal P}\,H=
 E_n\,^{(\vec{\kappa})}\!\langle E_n\,|\,{\cal P},
 \\
 ^{(\vec{\kappa})}\!\langle\!\langle E_m\,|\,H=
  E_m\,^{(\vec{\kappa})}\!\langle\!\langle E_m\,|.
 \end{gather*}
This would {\em change} the form of our proportionality rule
(\ref{psiv}), into
 \begin{gather*}
 |\,E_n\rangle\!\rangle^{(\vec{\kappa})}=
 {\cal P}^\dagger\,|\,E_n\rangle^{(\vec{\kappa})}\cdot
 c_n^{(\vec{\kappa})}
 \end{gather*}
with an immediate consequence
 \[
 1= ^{(\vec{\kappa})}\!\!\langle E_n\,|\,
 {\cal P}^\dagger
 \,|\,E_n\rangle^{(\vec{\kappa})}\cdot c_n^{(\vec{\kappa})}
 =^{(1)}\!\!\langle E_n\,|\,
 {\cal P}^\dagger\,|\,E_n\rangle^{(1)}\cdot
  c_n^{(\vec{\kappa})}\cdot
 \kappa_n^*\kappa_n
 \]
i.e.,
 \[
  c_n^{(\vec{\kappa})}= \frac{1}{\kappa_n^*\kappa_n}
  \,c_n^{(1)} , \qquad
  c_n^{(1)}=\frac{1}{^{(1)}\!\langle
 E_n\,|\,
 {\cal P}^\dagger\,|\,E_n\rangle^{(1)}}
 =
 \big (q_n^{(1)} \big)^*
  .
 \]
Now we may introduce the standard charge operator ${\cal C}={\cal
C}^{(\vec{\kappa})}$ by setting
 \begin{gather*}
 |\,E_n\rangle^{(\vec{\kappa})}
  \cdot c_n^{(\vec{\kappa})}=
  \big[{\cal C}^{(\vec{\kappa})}
  \big ]^\dagger\,|\,E_n\rangle^{(\vec{\kappa})}
 \end{gather*}
and
 \begin{gather*}
 \big [{\cal C}^{(\vec{\kappa})}
  \big ]^\dagger
 =\sum_{n=0}^\infty\,|\,E_n\rangle^{(\vec{\kappa})}
 \,{c_n^{(\vec{\kappa})}}\,
 ^{(\vec{\kappa})}\!\langle\!\langle E_n\,|=
 \sum_{n=0}^\infty\,|\,E_n\rangle^{(1)}
 \,\frac{c_n^{(1)}}{\kappa_n^*\kappa_n}\,
 ^{(1)}\!\langle\!\langle E_n\,|
  \end{gather*}
i.e.,
 \begin{gather*}
 {\cal C}^{(\vec{\kappa})}
  =
 \sum_{n=0}^\infty\,|\,E_n\rangle\!\rangle^{(1)}
 \,\frac{q_n^{(1)}}{\kappa_n^*\kappa_n}\,
 ^{(1)}\!\langle E_n\,| .
  \end{gather*}
This leads to a {\em non-equivalent} factorization
 \begin{gather*}
 \Theta^{(\vec{\kappa})}={\cal C}^{(\vec{\kappa})}{\cal P}
 = \sum_{n=0}^\infty\,|\,E_n\rangle\!\rangle^{(1)}
 \,\frac{q_n^{(1)}}{\kappa_n^*\kappa_n}\,
 ^{(1)}\!\langle E_n\,|\,{\cal P}  \nonumber \\
 \phantom{\Theta^{(\vec{\kappa})}}{} =\sum_{n=0}^\infty\,|\,E_n
 \rangle\!\rangle^{(1)}
 \,\frac{1}{\kappa_n^*\kappa_n}\,
 ^{(1)}\!\langle\!\langle E_n\,|=\sum_{n=0}^\infty\,|\,E_n
 \rangle\!\rangle^{((\vec{\kappa}))}
 \,\cdot\,
 ^{((\vec{\kappa}))}\!\langle\!\langle E_n\,|
  \end{gather*}
to be compared with formula (\ref{sosonek}).

\section{Summary}

Carl Bender \cite{Carl} lists several reasons why the usual
Hermiticity of the quantum Hamiltonians~$H$ (i.e., their property
$H = H^\dagger$ where the superscript symbolizes the matrix
transposition plus complex conjugation) should be replaced by the
better motivated rule~(\ref{ptsymm}). Although the latter relation
of\/fers just a typical sample of a ${\cal P}$-pseudo-Hermiticity of
$H$, it is often called, in the context of some older work in this
direction~\cite{BG}, ``${\cal PT}$-symmetry'' of $H$.

In this context, Mostafazadeh \cite{ali} noticed that on a purely
formal level, the symbol ${\cal P}$ need not coincide with parity
at all. He suggested and promoted its ``pseudometric''
reinterpretation preserving the Hermiticity ${\cal P}={\cal
P}^\dagger$ but relaxing the involutivity, ${\cal P} \neq {\cal
P}^{-1}$. The f\/irst step towards generalizations has been made.

Originally \cite{BB} it has been believed that the ${\cal
PT}$-symmetry of $H$ could possess a deeper physical signif\/icance,
especially when the operators ${\cal P}$ and ${\cal T}$ were
chosen as representing the physical parity and the time reversal,
respectively. Later on, it became clear that this property must be
{\em constructively} complemented by another, independent and much
more relevant antilinear symmetry (\ref{cptsymm}) called, mostly,
${\cal CPT}$-symmetry of $H$ (where ${\cal C}$ is called
``charge''). In the light of~\cite{ali} and \cite{Geyer},
just an expectable return to the safe waters of standard quantum
mechanics has been accomplished.

In the next step of development, Solombrino \cite{Solombrino} and
others \cite{str,BQ} admitted all  ${\cal P} \neq {\cal
P}^\dagger$ which remain invertible. In a way  complementing, and
inspired by, the related recent remark by Ali~Mostafazadeh
\cite{aliweak} we have shown here that after transition to
non-Hermitian auxiliary operators ${\cal P}\neq {\cal P}^\dagger$
the concept of charge (def\/ined as a pre-factor in the metric
$\Theta= {\cal CP}$) becomes ambiguous (in the sense that we could
also have $\Theta ={\cal C}'{\cal P}^\dagger$ in principle). We
clarif\/ied this ``puzzle'' by showing that there really exists {\em
another} auxiliary family of operators ${\cal Q}$ such that
$\Theta = {\cal P Q}={\cal Q}^\dagger{\cal P}^\dagger$. We may
note that the unavoidable Hermiticity of the metric $\Theta$
implies that the {\em weak} form of pseudo-Hermiticity leads to
the {\em richer} menu of the alternative forms of the
factorization of the metric,
 \[
 \Theta^{(\vec{\kappa})}
 ={\cal P Q}^{(\vec{\kappa})}
 ={\cal C}^{(\vec{\kappa})}\,{\cal P}
 =
 \big [\Theta^{(\vec{\kappa})}
 \big]^\dagger
 =
 \big [{\cal  Q}^{(\vec{\kappa})}\big ]^\dagger\,
 {\cal P}^\dagger
 = {\cal P}^\dagger\,
 \big[{\cal C}^{(\vec{\kappa})}\big ]^\dagger
.
\]
In the other words, whenever we relax the ``usual'' constraint
${\cal P} = {\cal P}^\dagger$, there emerges a certain
complementarity between the concepts of the charge ${\cal C}$ and
quasiparity ${\cal Q}$.

It is possible to summarize that on the present level of
understanding of the use of $H\neq H^\dagger$ in quantum
mechanics, people are aware that in the most relevant cases (when
the spectrum~$\{E_n\}$ of our~$H$ is all real, discrete and, for
the sake of brevity of formulae, non-degenerate), the role of
${\cal P}$ remains purely auxiliary. Still, a distinct boundary
between the ``feasible'' and ``not feasible'' applications seems to
coincide, more or less, precisely with the boundary between the
``suf\/f\/iciently simple'' and ``not suf\/f\/iciently simple'' operators
${\cal P}$ in~(\ref{ptsymm}). For this reason we tried here to
draw a few consequences from the use of some ``extremely simple"
${\cal P} \neq {\cal P}^\dagger$. We demonstrated that in both the
constructions of the bases and in the factorizations of the metric~$\Theta$ the use of the non-Hermitian ${\cal P}$  could have its
merits. Last but not least we also proved that the ``natural''
requirement ${\cal Q}^2=I$ of the involutivity of the quasiparity
(ref\/lecting its usual role in some applications
\cite{pseudo,ptho}) is fully equivalent to the more standard
recipe ${\cal C}^2=I$ which proved, in many models \cite{Carl}, so
useful for an ef\/f\/icient suppression of the well known \cite{Geyer}
enormous ambiguity of the metric $\Theta$.

\subsection*{Acknowledgement}

Work supported by GA\v{C}R, grant Nr. 202/07/1307, Institutional
Research Plan AV0Z10480505 and by the M\v{S}MT ``Doppler
Institute'' project Nr. LC06002.

\pdfbookmark[1]{References}{ref}
\LastPageEnding

\end{document}